\documentclass[twocolumn,preprintnumbers,,aip,apl]{revtex4}

\usepackage{graphicx,amsmath,amssymb}

\begin{document}

\title{Negative photoconductivity and hot-carrier bolometric
detection of  terahertz radiation  in  graphene-phosphorene hybrid structures
}
\author{V. Ryzhii$^{1,2,3,4*}$,  M. Ryzhii$^{5}$, D. S. Ponomarev$^{2,3}$,
V. G. Leiman$^{3}$, V. Mitin$^{1,6}$,
 M. S. Shur$^{7,8}$, and   T.~Otsuji$^{1}$}
\affiliation{
$^{1}$ Research Institute of Electrical Communication, Tohoku University,
 Sendai 980-8577, Japan\\
$^{2}$ Institute of Ultra High Frequency Semiconductor Electronics of RAS,
 Moscow 117105, Russia\\
$^{3}$  Center of Photonics and Two-Dimensional Materials, Moscow Institute of Physics and Technology, Dolgoprudny 141700, Russia\\
$^{4}$ Center for Photonics and Infrared Engineering, Bauman Moscow State Technical University,\\
  Moscow 111005, Russia\\
$^{5}$ Department of Computer Science and Engineering, University of Aizu,\\ 
Aizu-Wakamatsu 965-8580, Japan\\
$^{6}$ Department of Electrical Engineering, University at Buffalo, Buffalo, New York 1460-192\\
$^{7}$ Department of Electrical, Computer, and Systems Engineering
and Department of Physics, Applied Physics, and Astronomy, Rensselaer Polytechnic Institute, Troy, New York 12180, USA\\ 
$^{8}$
Electronics of the Future, Inc., Vienna, VA 22181, USA}
%

\begin{abstract}
\normalsize 
We consider the effect of  terahertz (THz) radiation on the conductivity
of the ungated and gated graphene (G)-phosphorene (P) hybrid structures and propose and evaluated the hot-carrier uncooled bolometric photodetectors based  on the  GP-lateral diodes (GP-LDs) and GP-field-effect transistors (GP-FETs) with
the GP channel.
 The  operation  of the GP-LDs and GP-FET  photodetectors is associated with the carrier heating by the  incident radiation absorbed in the G-layer due to the intraband transitions.
The carrier heating leads to the relocation of a significant fraction of the carriers into the P-layer.
Due to a  relatively low mobility of the carriers in the P-layer, their main role is associated with
a substantial reinforcement of the scattering of the carriers. The GP-FET bolometric photodetector characteristics are effectively controlled by the gate voltage. 
A strong
negative conductivity of the GP-channel can  provide much higher responsivity of the  THz  hot-carriers GP-LD and GP-FET bolometric photodetectors
in comparison with the bolometers with solely the G-channels. 
\end{abstract}

\maketitle
\newpage


 \section{Introduction}  

Unique energy spectra, of  graphene (G)~\cite{1} 
and a few-layer Black Phosphorus layer or phosphorene (P)~\cite{2},  their
optical and electric properties, and
 recent advances in technology open  remarkable prospects for the creation of novel devices using
G-layers~\cite{3,4,5,6}, the P-layers~\cite{2,7,8,9,10},  and different hybrid structures including the  G-P  hybrid structures~\cite{11,12,13}. In particular, the GP hybrid systems  can be used for the improvement of various devices. The possibility  of the layer-dependent alignment work function control~\cite{2,14}  provides substantial 
flexibility in the device design. 
 In this paper, we propose and evaluate the detector of the terahertz (THz) radiation based on a lateral diode  (LD) and a
 field-effect transistor (FET) with the GP channel, GP-LD and GP-FET, respectively.
 The  operation  of the GP-LD and GP-FET photodetectors  with the GP-channel 
 is associated with the carrier heating by the  incident radiation absorbed in the G-layer leading to a variation
 of the channel conductivity~\cite{15,16}. This principle is used in the hot-carrier bolometers based on the G-channel exhibiting 
 the negative or positive photoconductivity (see, for example~\cite{17,18,19,20,21} and references therein).
 However,  a major disadvantage of using G-layers in the bolometric photodetectors is that  the conductivity of pristine
 G-layers is weakly dependent on the carrier temperature. This can be overcome by the introduction of the barrier regions (by partitioning of the channel into  nanoribbons in which the energy gap is opened~\cite{17} or using disordered G-layers~\cite{20}).
 In the G-P bolometers under consideration,
 the carrier heating caused by the absorbed radiation  leads to the transfer of a significant portion of the carriers into the P-layer. This results in a decrease of the density of the highly mobile carriers in the G-layer and in a reinforcement of the scattering of these carriers on the carriers residing in the P-layer. 
Due to a high effective mass of the carriers in the P-layer and, hence, a  relatively low mobility, their main role is associated with
a substantial reinforcement of the scattering of the carriers in the G-layer. 
As a result, the conductivity of the GP-channel can markedly drop with the carrier heating.
We demonstrate that the effect of the negative THz photoconductivity  in the G-P channels can be much stronger than that for the G-channels, particularly at  room temperature.
Therefore, the GP-LDs and GP-FETs could effectively operate as the uncooled hot-carrier  THz bolometers
with an elevated responsivity.

\section{Model}

\begin{figure*}[t]
\centering
\includegraphics[width=11.0cm]{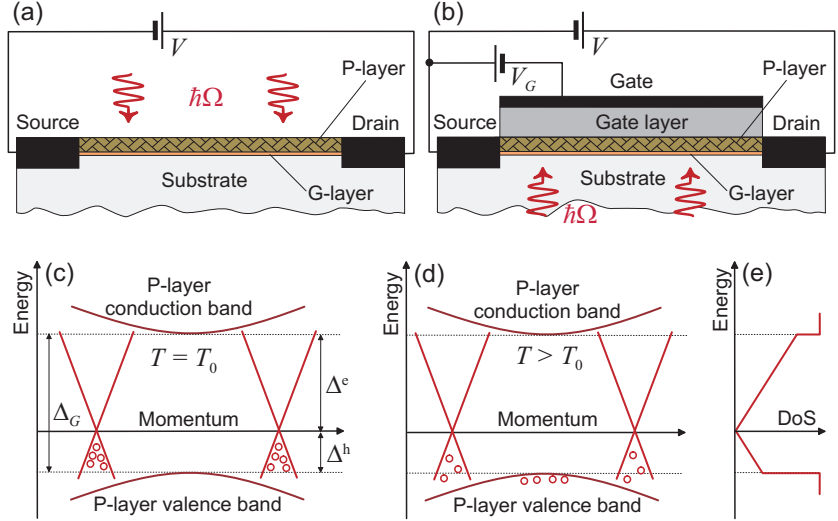}
\caption{The structures of  (a) the GP-LD  and (b) the GP-FET (b),  their asymmetric 
with respect to the Dirac point ($\Delta^e > \Delta^h$) energy band diagrams with the G-Dirac cones and the parabolic extrema corresponding to the P-layer at (c)  $T = T_0$ and (d)   $T > T_0$,  
and  (e) the energy dependence  of the  density of state (DoS).
 Open circles correspond to the holes
in the valence bands of G- and P-layers.
}
\label{F1}
\end{figure*}
  
  Figure~1 demonstrates a schematic view of the ungated and gated G-P structures (i.e., a GP-LD and a GP-FET  with the GP-channels). It is assumed that the P-layer  consisting of a few atomic layers, is oriented in such a way that the direction from the source to its drain  corresponds to the zigzag direction.
The dynamics of electrons and holes in this direction is characterized by a huge effective mass.
As a result, a substantial amount  of the electrons and holes can relocate   from the G-layer (where their mobility could be very high) to the  P-layer (with a low mobility).

 For the sake of definiteness, we consider the P-layer consisting of several atomic P-layers ($N= 4-5$), assuming that in both GP-LD and GP-FET structures it is  p-doped (the pristine P-layers are of p-type). 
At the incident THz photon energies $\hbar\Omega < 2\mu$, where $|\mu|$ is the Fermi energy of the main carriers (holes), the carrier heating
is associated with its interband (Drude) 
 absorption in the G-layer.

 The band gap $\Delta_G$ and the energy spacing $\Delta^e$ and $\Delta^h$,
between the Dirac point in the G-layer and the edges 
of the conduction and valence bands,  (determined by the pertinent work functions) depend on the number $N$.
In  the G-P channel under consideration (with $N = 4$ or 5), the band structure is asymmetric: $\Delta^e > \Delta^h = \Delta$~\cite{14}.

The dispersion relation for the  holes in the G- and P-layers can be presented as

\begin{equation}\label{eq1}
 \varepsilon_G^e = v_W\sqrt{p_x^2 + p_y^2}, \qquad  \varepsilon_P^e = \Delta^e + \frac{p_x^2}{2m_{xx}} + 
\frac{p_y^2}{2m_{yy}},
\end{equation}

\begin{equation}\label{eq2}
 \varepsilon_G^h = -v_W\sqrt{p_x^2 + p_y^2}, \qquad  \varepsilon_P^e = - \Delta^h - \frac{p_x^2}{2m_{xx}}  
-\frac{p_y^2}{2m_{yy}},
\end{equation}
respectively. Here $v_W \simeq 10^8$~cm/s  is the characteristic velocity of electrons in the G-layers,
$m_{xx} = M$ and $m_{yy} =m$ are the components of the effective mass tensor ($M \gg m$), $p_x$ and $p_y$
are the carrier momenta in the source-drain direction and the perpendicular direction, respectively.
The components of the effective mass tensor for the valence bands ($N = 4 - 5$) are approximately equal to
$m  \simeq 0. 04m_0$ and $M \simeq 1.01m_0$, where $m_0$ is the mass of a free electron.

The model under consideration is based on the following assumptions:

(1) The sufficiently frequent electron-electron, electron-hole, and  hole-hole collisions enable the establishment of the distinct quasi-Fermi electron and hole distributions with 
the common effective temperature $T$  in both G- and P-layers
and  split  quasi-Fermi levels due to the carrier-phonon interband scattering. 
This is consistent with the numerous experimental studies in which the G-layer was excited with an optical or infrared pump pulse and probed with photoelectron or optical spectroscopy at different photon energies (see, for example,~\cite{18} and the references therein).
Sufficiently strong  interactions between the electrons and 
holes belonging to both G- and P-layers promote the inter-layer equilibrium~\cite{22,23,24}.  
Hence,
 the electron and hole distribution functions are the following functions of the 
carrier energy $\varepsilon$: $f^{e,h} = [\exp(\varepsilon - \mu^{e,h})/T +1]^{-1}$
 (where $\mu^e$ and $\mu^h = \mu$ are the quasi-Fermi energies counted from the Dirac point).

(2) Due to heavy electron and hole effective masses $M$ and $\sqrt{mM}$, 
the conductivity of the P-layer is relatively small because this layer mobility in the direction corresponding to the mass $M$ is proportional to $ 1/\sqrt{mM}M$~\cite{25}, so that 
 the P-layer conductivity could be neglected  in comparison with
the G-layer conductivity. Thus, the main role of the carrier relocation from the G-layer into the P-layer
is associated with an intensification of the  carriers (in the G-layer) scattering on the  carriers (in the P-layer) when the concentration of the latter increases with the carrier system heating.\\

(3) The  momentum relaxation of the electrons and holes in the G-layer (which we refer to as the "light" electrons and holes) is due to their scattering on  acoustic phonons, neutral defects, and  heavy particles in the P-layer. In contrast to the G-channel-based THz bolometers intended for the operation at very low temperatures at which the carrier energy relaxation is due to the interaction with  acoustic phonons,   
the energy relaxation in the uncooled bolometric detectors under consideration is associated with the optical phonons in the G-layer.
The interband transitions assisted by the optical phonons~\cite{26,27} and with  the Auger processes (see ~\cite{28,29} and the discussion therein)are assumed to be  the  main recombination-generation mechanisms.
We characterize the relative role of these processes by the parameter $\eta = \tau_{Auger}/(\tau_{Auger}+\tau_{0}^{inter})$, where $\tau_{Auger}$ and $\tau_{0}^{inter}$ are the times characterizing the pertinent interband transitions (we call this parameter as the Auger parameter). When $\eta \simeq 1$, the quasi-Fermi energies can be markedly different  ($\mu^e \neq -\mu^h$). 

\section{Conductivity of the G-P-channel}

The net surface charge  density in the GP-channel, which comprises the electron and hole charges in both the G- and P-layers,  induced by  the acceptors  and the gate voltage $V_G$  is equal to    $e\Sigma = \kappa\,|V_G - V_{A}|/4\pi\,W_g$ (where $V_G =  V_{A} > 0$, which is proportional to the acceptor density,  corresponds to the charge neutrality point,  $\kappa$ and $W_g$ are the background dielectric constant and the thickness of the gate layer, and $e$ is the electron charge). By introducing the voltage (gate) swing $V_g = V_G - V_A$, we unify  the consideration of the GP-LDs and GP-FETs. In particular, the case of GP-LDs corresponds to $V_G = 0$, so that $V_g = - V_A < 0$, while in the GP-FETs $V_g$ can be both negative and positive. 

The gate voltage swing $V_g = V_G - V_{A}$ or its dimensional value $U_g = V_g/V_0$ and the quantities $T$, $\mu^e$, and $\mu^h$ are related to each other as
\begin{eqnarray}\label{eq3}
 U_g= \biggl(\frac{T}{T_0}\biggr)^2 \biggl[{\cal F}_{1}\biggl(\frac{\mu^e}{T}\biggr)
-  {\cal F}_{1}\biggl(\frac{\mu^h}{T}\biggr)\biggr]\nonumber\\
-
\gamma_N\frac{T}{T_0}\ln
\biggl[1 +\displaystyle\exp\biggl(\frac{\mu^h -\Delta}{T}\biggr)\biggr].
\end{eqnarray}
Here  
${\cal F}_1(a)=\int_0^{\infty}d\xi\xi[\exp(\xi - a) + 1]^{-1}$
is the Fermi-Dirac integral~\cite{30}, $U_g = (V_G - V_A)
/V_0 = V_G/V_0 - \pi\Sigma_A\hbar^2v_W^2/2T_0^2/$,
$V_0 = 8eT_0^2W_g/\kappa\hbar^2v_W^2$, 
$\gamma_N = N\sqrt{mM}v_W^2/2T_0$,   and  $\hbar $ is the Planck constant.
For $N = 5$ and therefore setting ,  ($\sqrt{mM} \simeq 0.2m_0$
and $M/m \simeq 25$),  $W_g = 10 - 1000$~nm and $\kappa = 4$, and $T = 25$~meV ($\simeq 300$~K), one can obtain  $\gamma_N \simeq 110$ and $V_0 \simeq 0.05 -  4.94$~V. A large value of $\gamma_N$ is due to a relatively high density of states
in the P-layer.

In equilibrium at sufficiently low  temperatures, when the P-layer is empty (the second term in the right-hand side of  Eq.~(3) is negligible), Eq.~(3) yields $\mu^e = -\mu^h = -\mu \simeq -\hbar\,v_W\sqrt{\kappa|V_g|/4eW_g}$ when $|V_g|$ is relatively large.

When the electron-hole system in the G-P channel is heated by the source-to-drain DC voltage or by the incident
radiation, the electron and hole quasi-Fermi levels can be split: $\mu^e \neq - \mu^h$. 
Accounting for the competition between the optical phonon mediated and the Auger
 generation-recombination processes, the equation governing the carrier interband balance can result in the following equation relating $\mu^e$ and $\mu^h$ at an arbitrary effective temperature $T$:

\begin{equation}\label{eq4}
\mu^e + \mu^h = \eta\hbar\omega_0\biggl(1 - \frac{T}{T_0}\biggr).
\end{equation}
where $\hbar\omega_0$ is the optical phonon energy. Equation~(4) generalizes that obtained previously~\cite{13} for the case
of the dominant optical phonon generation-recombination processes  by the introduction of a phenomenological  factor $\eta = \tau_{Auger}/(\tau_{Auger} + \tau_{Opt})$.

Considering~ Eq.~(4), we rewrite Eq.~(3) as
\begin{eqnarray}\label{eq5}
 U_g = \biggl(\frac{T}{T_0}\biggr)^2 \biggl[{\cal F}_{1}\biggl(-\frac{\mu}{T} - \eta\hbar\omega_0\biggl(\frac{1}{T_0} - \frac{1}{T}\biggr)\biggr)
-  {\cal F}_{1}\biggl(\frac{\mu}{T}\biggr)\biggr]\nonumber\\
-
\gamma_N \frac{T}{T_0}\ln
\biggl[1 +\displaystyle\exp\biggl(\frac{\mu -\Delta}{T}\biggr)\biggr].
\end{eqnarray}
In particular, using Eq.~(5), one can obtain immediately the dependence of the hole Fermi energy $\mu_0 = \mu|_{T= T_0}$ on 
the voltage swing $U_g$.

Focusing on the GP-channels with dominant carrier  scattering on acoustic phonons, neutral defects, on each other, and 
on the short-range screened heavy carriers, the momentum relaxation time $\tau(p)$ as a function of the carrier momentum $p$ can be set as 
$\tau_p = \tau_0(T_0/pv_W)[\Sigma_G/(\Sigma_G + \Sigma_P)]$, where $\tau_0 \propto \Sigma_G^{-1}$ is the momentum relaxation time in the G-layer with the effective scatterer density $\Sigma_G$  at $T = T_0$ and $(\Sigma_G + \Sigma_P)$ is the net scatterer density, which 
accounts for  the density, $\Sigma_P$, of the heavy carriers in the P-layer.
In this case,  the GP-channel conductivity could be presented as  (in line with~\cite{15,16,31,32,33,34,35,36}):

\begin{equation}\label{eq6}
 \sigma_{GP} = -\frac{\sigma_0\Sigma_G}{(\Sigma_G + \Sigma_P)} 
\int_0^{\infty}d\xi\frac{d(f^e + f^h)}{d\xi}
\end{equation}
with   $\sigma_{0} =  (e^2T_0\tau_{0}/\pi\hbar^2)$  being   the G-layer low electric-field conductivity.
Using Eqs.~(4) and (6), the GP-channel conductivity can be expressed via the G-layer conductivity $\sigma_{G}$ (without the P-layer conductivity) with the latter expressed via the effective temperature $T$
and the hole quasi-Fermi energy $\mu = \mu^h$:

\begin{equation}\label{eq7}
 \sigma_{GP} = \frac{\sigma_G\Sigma_G}{(\Sigma_G + \Sigma_{P})},
\end{equation}

\begin{eqnarray}\label{Eq8}
\sigma_G = \sigma_0\biggl[
 \frac{1}{\displaystyle\exp\biggl(\frac{\mu}{T} + \eta\hbar\omega_0\biggl(\frac{1}{T_0} - \frac{1}{T}\biggr)\biggr) + 1}\nonumber\\
  + \frac{1}{\displaystyle\exp\biggl(-\frac{\mu}{T}\biggr) + 1}  \biggr].
\end{eqnarray}

The  density of scatterers (heavy holes), $\Sigma_P$, in the P-layer, which   exponentially increases with increasing $|\mu|$ and $T$, can also be expressed via these quantities: 
\begin{eqnarray}\label{eq9}
 \Sigma_{P}  
= \Sigma_N
\frac{T}{T_0}\displaystyle\ln\biggl[1 +\exp\biggl(\frac{\mu - \Delta}{T} \biggr)\biggr],
\end{eqnarray}
 where $\Sigma_N = N T_0\sqrt{mM}/\pi\hbar^2 \propto \gamma_N$. The factor $N$ in the latter formula reflects the fact that the density of states in the few-layer P-layer increases  with the layer number $N$~\cite{15}.

The second factor in the right-hand side of Eq.~(6)  reflects an increase in the scatterer density  associated with the inclusion of the scattering on the heavy holes in the P-layer.
It is instructive that at $\eta = 0$ when $\mu^e = -\mu^h = -\mu$, the G-channel conductivity $\sigma_G= \sigma_0$ is independent of $T$. This is because of the specific of the carrier scattering in the system under consideration (scattering on acoustic phonons, neutral defects and effectively screened charged scatterers)~\cite{35,36,37,38}.

Using Eqs.~(7) - (9), we obtain

\begin{eqnarray}\label{eq10}
 \sigma_{GP} = \frac{\sigma_G}{1 + \displaystyle \frac{\Sigma_{N}}{\Sigma_G}\frac{T}{T_0}\ln\biggl[1 +\exp\biggl(\frac{\mu - \Delta}{T}\biggr)\biggr]}\nonumber\\
=\frac{\sigma_0}{1 + \displaystyle \frac{\Sigma_{N}}{\Sigma_G}\frac{T}{T_0}\ln\biggl[1 +\exp\biggl(\frac{\mu - \Delta}{T}\biggr)\biggr]}\nonumber\\
\times\biggl[
 \frac{1}{\displaystyle\exp\biggl(\frac{\mu}{T} + \eta\hbar\omega_0\biggl(\frac{1}{T_0} - \frac{1}{T}\biggr)\biggr) + 1} + \frac{1}{\displaystyle\exp\biggl(-\frac{\mu}{T}\biggr) + 1}  \biggr]. 
\end{eqnarray}
Equations~(5) and (10) describe the dependences of the quasi-Fermi energy $\mu$ 
and   the G-P-channel conductivity $\sigma_{GP}$  on the effective temperature $T$ and the  voltage swing $U_g$.
Solving these equations,
one can obtain the characteristics of the GP-channel 
in wide ranges of  the normalized voltage swing $U_g$, carrier effective temperature $T$, and the  density $\Sigma_G$.

\section{Negative  photoconductivity in the G-P channels}

\begin{figure*}[t]
\centering
\includegraphics[height=5.0cm]{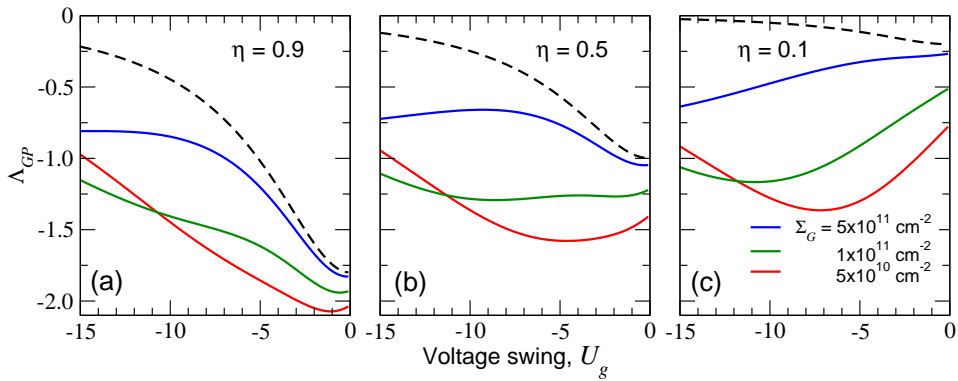}
\caption{The temperature derivative of the  GP-channel conductivity, $\Lambda_{GP}$,  versus $U_g$ for $\Delta = 200$~meV  and different scatterer densities $\Sigma_G$  
at  (a) $\eta = 0.9$,  (b) $\eta  = 0.5$, and (c) $\eta = 0.1$. Dashed line corresponds to 
$\Lambda_G$  ($N = 0$,i.e., $\Sigma_N =0$) and the same values of $\eta$.} 
\label{F2abc}
\end{figure*}

The variation of the current density, $J - J_0$, in the GP-channel for small  effective carrier temperature variations
is given by

\begin{equation}\label{eq11}
J - J_0 \simeq E_{SD}\frac{d\sigma_{GP}}{d T}\biggr|_{T = T_0}\cdot(T - T_0),
\end{equation}
where $J_0$ is the linear density of the source-drain dc  current in the absence of the irradiation (i.e., the dark current density, $E_{SD} = V_{SD}/L$ and $V_{SD}$ are the source-drain electric field and voltage, respectively, $L$ is the length of the GP-channel,  and $\sigma_{GP}$ is given by Eq.~(10).  An increase  in the carrier effective temperature  $T$ (the carrier heating) caused by  the irradiation corresponds to the negative photoconductivity temperature when the conductivity derivative
$(d\sigma_{GP}/d T)|_{T = T_0} < 0$. The latter  is in line with the experimental observations~\cite{18,19,37,38}.
Figure~2 shows $\Lambda_{GP} = \displaystyle\frac{1}{\sigma_0}\frac{d\sigma_{GP}}{d\ln T}\biggr|_{T = T_0}$ and $\Lambda_{G} = \displaystyle\frac{1}{\sigma_0}\frac{d\sigma_{G}}{d\ln T}\biggr|_{T = T_0}$ 
 (i.e., with $N = 0$)
found as  functions of 
$U_g$  using Eqs.~(5) and (10) for different values of the scatterer density $\Sigma_G$ in the G-layer.
One can see from Fig.~2 that both the quantities $\Lambda_{GP}$ and $\Lambda_G$
are negative. Here and in the following, we assume 
 that $\hbar\omega_0 = 200$~meV, $\Delta = 200$~meV,  $\Sigma_N = 1.2\times 10^{13}$~cm$^{-2}$, $\Sigma_G = 5\times(10^{10} - 10^{11})$~cm$^{-2}$ corresponding to 
 $\tau_0 = (0.24 - 2.4)$~ps,  $\eta = 0.1 - 0.9$,   $T_0 = 25$~meV ($\simeq$ 300K), and $\gamma_N = 110$.
 The scatterer densities range $\Sigma_G = 5\times(10^{10} - 10^{11})$~cm$^{-2}$
at $\mu_0 = 75$~meV, corresponds to the rather practical values of the carrier mobility in G-layers 
$b_G \simeq (30 - 300)\times 10^3$~cm$^2$/s V~\cite{39,40,41,42}.

At small $|U_g|$,  the absolute values of these quantities $|\Lambda_{GP}| \gtrsim |\Lambda_G|$.
However, 
at sufficiently large $|U_g|$,  $|\Lambda_{GP}|$ can be substantially larger
than $|\Lambda_G|$, particularly, at $\eta \ll 1$.
This is attributed to a steeper effective temperature dependence of the GP-channel  conductivity due
to an increasing hole population in the P-layer. The latter implies that the GP-channel can exhibit a stronger temperature dependence and, hence, 
a stronger effect of the negative photoconductivity than the G-channel.
The comparison of the plots in Figs.~2(a) -- 2(c) demonstrates that the relative intensification of the Auger processes (a decrease in the parameter $\eta$) leads to a marked decrease  in $|\Lambda_G|$,
diminishing the temperature dependence of the G-channel, while these processes weakly affect the temperature dependence of the GP-channel and, hence,  the quantities
 $|\Lambda_{GP}|$ and $J - J_0$.


\section{Responsivity  of the GP-photodetectors}

We limit our following consideration by the  GP photodetectors operating as hot-carrier bolometers, so that
the incident radiation does not produce a marked amount of the extra electrons and holes and the variation
of the carrier density is associated primarily with the heating processes. 
This happens when 
 $|U_g|$ is sufficiently large (to provide
large hole Fermi energy $\mu$), the photon energy $\hbar\Omega$ is not too large ($\hbar\Omega < 2\mu_0$). 
and the carrier momentum relaxation time $\tau_0$ is not too short, so that the intraband absorption dominates the interband absorption (see Appendix A).

Under irradiation, the carrier effective temperature varies. Its value can be found considering the balance of the power, $S_{abs}$, receiving by the carriers due to the absorption of the incident radiation with the photon energy $\hbar\Omega$
and the power, $S_{lattice}$, which the carriers transfer to the lattice. 
As assumed above, the latter is associated with
the interband transitions accompanied by the emission and absorption of the G-channel optical phonons
having the energy $\hbar\omega_0$.
The power received by the carrier system is given by 

\begin{equation}\label{eq12}
S_{abs} = \frac{4\pi\sigma_{GP}}{c\sqrt{\kappa}}\hbar\Omega\,I_{\Omega} = \frac{4\pi}{c\sqrt{\kappa}}\frac{\sigma_{GP}}{(1 + \Omega^2\tau^2)}\hbar\Omega\,I_{\Omega}.
\end{equation}
where $c$ is the speed of light, $I_{\Omega}$ is the radiation photon flux,
$\sigma_{GP,\Omega} =\sigma_{GP}/(1 + \Omega^2\tau^2)$ is the high-frequency G-P channel conductivity,
$\tau$ is the average hole momentum relaxation
time in the GP-channel, which considering that $\tau_p \propto 1/p $, can be estimated as

\begin{equation}\label{eq13}
\tau \simeq \frac{1}{[2[{\cal F}_{1}(\mu_0/T_0) + {\cal F}_{1}(-\mu_0/T_0)]^{1/2}}\frac{ \tau_0}{(1 +P_N)}.
\end{equation}
Here $P_N =(\Sigma_{N}/\Sigma_G)\ln\{1 + \exp[- (\Delta -\mu_0)/T_0]\}$.
At $\mu_0 \lesssim T_0$ and $\mu_0 \gg T_0$, Eq.~(13) yields 
$\tau \simeq [\sqrt{3}\tau_0/\pi(1 + P_N)]$ and $\tau \simeq [\tau_0/(1 + P_N)](T_0/\mu_0)]\simeq \tau_0/\sqrt{2U_g}$,
respectively.

Taking into account Eqs.~(7), (10), and (12) at $T = T_0$, we obtain

\begin{equation}\label{eq14}
S_{abs} = \biggl(\frac{4\pi\sigma_{0}}{c\sqrt{\kappa}}\biggr)\frac{\hbar\Omega\,I_{\Omega}}
{(1 + P_N)}\frac{1}{(1 + \Omega^2\tau^2)},
\end{equation}

\begin{figure*}[t]
\centering 
\includegraphics[height=5.0cm]{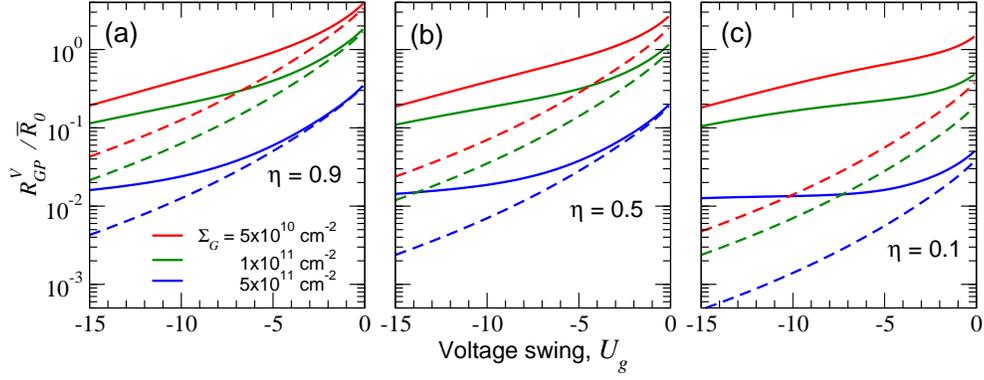}
\caption{The normalized GP-bolometer voltage responsivity $R_{GP}^V/\overline {R}_{0}^V$ versus the  voltage swing $U_g$ for the same parameters as in Fig.~2: (a) $\eta = 0.9$, (b) $\eta = 0.5$, and (b) $\eta = 0.1$.
 Dashed lines correspond to the G-bolometers with the same parameters of the G-layer.
}
\label{F4abc}
\end{figure*}

\begin{figure}[b]
\centering 
\includegraphics[height=5.0cm]{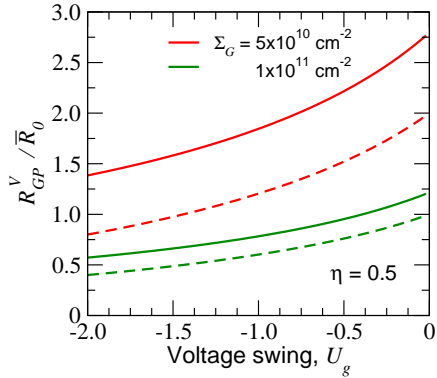}
\caption{Zoom of the same plots as in Fig.~3(b).}
\label{F5}
\end{figure}

As previously~\cite{17,27}, we set

\begin{equation}\label{eq15}
S_{lattice} = \hbar\omega_0\frac{\Sigma_0}{\tau_0^{intra}}\biggl[({\cal N}_{0} + 1) \exp\biggl(-\frac{\hbar\omega_0}{T}\biggr) - {\cal N}_0\biggr]
\end{equation}
where ${\cal N}_0 \simeq \exp(-\hbar\omega_0/T_0)$ is the equilibrium number of optical phonons,
$\tau_0^{inter}$ is the characteristic time of the spontaneous emission of optical phonons at the intraband transitions, and  $\Sigma_0 \simeq \pi\,T_0^2/3\hbar^2v_W^2$ (at $\mu_0 \lesssim T_0$ and $\Sigma_0 \simeq \mu_0^2/\pi\hbar^2v_W^2$
(at $\mu_0 \gg T_0) $ are the carrier densities in the G-channel at $T=T_0$. For simplicity, below we use
the following interpolation formulas:

\begin{equation}\label{eq16}
\tau = \frac{\tau_0}{(1 + P_N)}\sqrt{\frac{3}{\pi^2\displaystyle\biggl(1 + \frac{6|U_g|}{\pi^2}\biggr)}},
\end{equation}

\begin{equation}\label{eq17}
 \Sigma_0 = \frac{\pi\,T_0^2}{3\hbar^2v_W^2}\biggl(1 + \frac{6|U_g|}{\pi^2}\biggr).
\end{equation}

Equalizing $S_{abs}$ and $S_{lattice}$ given by Eqs.~(14) and (15), and taking into account Eq.~(16), we arrive at the following equation which relates the variation of the carrier effective temperature $T - T_0$ and the photon flux $I_{\Omega}$:

\begin{eqnarray}\label{eq18}
\frac{T - T_0}{T_0} = \biggl(\frac{12\alpha\hbar\,v_W^2\tau_0\tau_0^{\varepsilon}}{\pi\,\sqrt{\kappa}T_0^2}\biggr)\nonumber\\
\times\frac{
\hbar\Omega\,I_{\Omega}} {(1 + P_N)(1 + 6|U_g|/\pi^2)(1 + \Omega^2\tau^2)}, 
\end{eqnarray}
where $\alpha = e^2/c\hbar \simeq 1/137$ is the fine structure constant.
 The latter formula corresponds to the hole energy relaxation time 
$\tau_0^{\varepsilon} = \tau_0^{intra}(T_0/\hbar\omega_0)^2\exp(\hbar\omega_0/T_0)$.
Setting $\tau_0^{intra} = 0.7$~ps (for example,~\cite{43}), one obtains $\tau_0^{\varepsilon} \simeq 32.6$~ps.
%

Taking into account  the variation of the current density, $J - J_0$, in the GP-channel  caused by the irradiation,
the GP bolometer intrinsic current responsivity $R_{GP}$
can be presented by  the following expression:

\begin{equation}\label{eq19}
R_{GP} = \frac{(J-J_0)H}{\hbar\Omega\,I_{\Omega}A}.
\end{equation}
Here   $A = LH$ is the GP-channel area, $L$ is the channel length (the spacing between the source and the drain), and
$H$  is the channel width, i.e., its   size in the direction perpendicular to the current  direction.
Using Eqs.~(11) and (18), we arrive at the following:

\begin{equation}\label{eq20}
R_{GP}  = R_0 \frac{|\Lambda_{GP}|
} {[(1 + P_N)(1 + 6|U_g|/\pi^2)(1 + \Omega^2\tau^2)]}
\end{equation}
Here
\begin{equation}\label{eq21}
R_0  = \frac{12\alpha}{\pi^2\sqrt{\kappa}} \frac{e}{T_0}\frac{ev_W^2\tau_0^2\tau_0^{\varepsilon}E_{SD}}{\hbar L}
\propto \tau_0^2,
\end{equation}

In the above calculations we have not accounted for the carrier heating by the source-drain electric field assuming that it is weak, so that $T$ is very close to $T_0$.
The pertinent condition is as follows:

\begin{eqnarray}\label{eq22}
E_{SD} \ll \overline{ E}_{SD}
= \frac{\pi\, T_0}{ev_W}\sqrt{\frac{(1  + P_N)(1 + 6|U_g|/\pi^2)}{3\tau_0\tau_0^{\varepsilon}}}\nonumber\\
 = \overline{E^*}_{SD}\sqrt{\frac{(1  + P_N)(1 + 6|U_g|/\pi^2)}{3}}.
\end{eqnarray}

For $E_{SD} = \overline{E^*}_{SD} = (\pi\,T_0/ev_W\sqrt{\tau_0\tau_0^{\varepsilon}})$, the  quantity $R_0$, given by Eq.~(21), is equal to

\begin{equation}\label{eq23}
{\rm max }R_0  = \frac{12\alpha}{\pi\sqrt{\kappa}} \frac{v_W\tau_0^{3/2}\sqrt{\tau_0^{\varepsilon}}}{\hbar L}
\propto \tau_0^{3/2}.
\end{equation}
At $\tau_0 = (0.24 - 2.4)$~ps,  $\tau_0^{\varepsilon} = 32.6$~ps, and $L = 10^{-3}$~cm, for the quantities  $\overline{E}_{SD}^*$ and  max~$R_0$ one can find $\overline {E}_{SD}^* \simeq (92 - 290)$~V/cm and   max~$R_0 \simeq (1.5 -43.3)$~A/W, respectively.

The voltage responsivity $R_{GP}^V = R r_L $, where $r_L$ is the load resistance. Setting $r_L$ equal to the GP-channel resistance, i.e., $r_L = L(1 + P_N)/H\sigma_0$. 
In this case, for $R_{GP}^V$ one obtains 

\begin{equation}\label{eq24}
R_{GP}^V  = R_0^V \frac{|\Lambda_{GP}|} {[(1 + 6|U_g|/\pi^2)(1 + \Omega^2\tau^2)]}
\end{equation}
with

\begin{equation}\label{eq25}
R_0^V  = \frac{12\alpha}{\pi\sqrt{\kappa}}\frac{\hbar\,v_W^2\tau_0\tau_0^{\varepsilon}E_{SD}}{T_0^2H} \propto \tau_0.
\end{equation}

Setting $E_{SD} = \overline{E^*}_{SD}$, we arrive at the following
expression for the characteristic value of the GP bolometer voltage responsivity:

\begin{equation}\label{eq26}
{\rm max} R_0^V   = \frac{12\alpha}{\sqrt{\kappa}}\sqrt{\tau_0\tau_0^{\varepsilon}}\biggl(\frac{\hbar\,v_W}{eT_0H}\biggr). 
\end{equation}
For $\tau_0 = 0.24 - 2.4$~ps, $\tau_0^{\varepsilon} = 32.6$~ps, and $H = 10^{-2}$~cm,
we obtain ${\rm max} R_0^V \simeq (1.85 - 5.85)\times 10^2$~V/W. Naturally, at weaker source-drain electric fields $E_{SD} < \overline {E^*}_{SD} \sim \overline {E}_{SD}$, 
the quantities $R_0$ and $R_0^V$ are  smaller than max$R_0$ and   max$R_0^V$.

Using  Eqs.~(20) and (24) at relatively low frequencies $\Omega$, the current and voltage responsivities  can be presented as

\begin{equation}\label{eq27}
R_{GP}  = \frac{ R_0|\Lambda_{GP}|} {(1 + P_N)(1 + 6|U_g|/\pi^2)},\,
R_{GP}^V  =  \frac{ R_0^V|\Lambda_{GP}|} {(1 + 6|U_g|/\pi^2)}.
\end{equation}

Figures~3 and 4 show the GP-bolometers voltage low-frequency responsivity $R_{GP}^V$ normalized by the quantity
$\overline {R}_{0}^V = R_0^V\biggr|_{\tau_0= 1.2ps}$ as a function of the voltage swing $U_g$ calculated
using Eqs.~(10) and (27) for  the same  structural parameters as for Fig.~3 (given in Sec.~4).
 The normalized responsivity of the G-detectors (with the G-channel) is also shown by the dashed lines.
First, as seen from Figs.~3 and 4, the responsivity sharply decreases with an increase in the scatterer density $\Sigma_G$ and the
voltage swing $U_g$. This is attributed to a weaker carrier heating at their stronger scattering   and their larger
density. The latter markedly rises with increasing $U_g$. Second,  the GP-bolometer  responsivity, 
  moderately exceeding that of the G-bolometer responsivity at small $|U_g|$, becomes orders of magnitudes  
larger at elevated values of $|U_g|$ (compare the solid and dashed lines in Figs.~3 and 4).
The difference in the  GP- and G-bolometers responsivities becomes fairly pronounced 
at smaller   Auger parameter $\eta$ (at stronger
Auger generation-recombination processes). This correlates with a drop of $\Lambda_G$ clearly seen in Fig.~2.

\section{Bandwidth and gain-bandwidth product}

As follows from Eqs.~(20) and (24), the GP-bolometer responsivity  decreases when the photon frequency $\Omega >  2\pi\,f_{GP}$, where the cut-off frequency is given by

\begin{eqnarray}\label{eq28}
f_{GP} = \frac{(1+P_N)}{2\sqrt{3}\tau_0}\sqrt{\biggl(1 + \frac{6|U_g|}{\pi^2}\biggr)}\nonumber\\
\simeq \frac{1}{2\sqrt{3}\tau_0}
\biggl[1+\displaystyle\biggl(\frac{\Sigma_N}{\Sigma_G}\biggr)\exp\biggl(\sqrt{2U_g} - \frac{\Delta}{T_0}\biggr)\biggr]\nonumber\\
\times\sqrt{\biggl(1 + \frac{6|U_g|}{\pi^2}\biggr)}.
\end{eqnarray}
At small $|U_g|$, Eq.~(27) yields $f_{GP} \simeq (1/2\sqrt{3\pi}\tau_0)$, so  that for the values of $\tau_0$
used above, one obtains   $f_{GP} \simeq 0.12 - 1.2$~THz. At relatively large $|U_g|$, the frequency $f_{GP} \simeq (1+P_N)\sqrt{2|U_g|}/2\pi$
can be much higher than that at $U_g \simeq 0$. This is seen from  
Fig.~5, which  shows the cut-off frequency $f_{GB}$ as a function of the normalized voltage swing calculated
using Eq.~(28). The cut-off frequency $f_{GP}$ is larger than the pertinent frequency for the G-bolometers $f_{G}$ by  a factor 
$(1 + P_N)$. At large values of $\Sigma_N/\Sigma_G$ and $U_g$, this factor can be larger than that at $U_g \simeq 0$.

The comparison of the 
 gain-bandwidth  products of the GP- and G-bolometers, defined as max~$R_{GP}^V f_{GP}$
 and $R_{G}^Vf_{G}$, yields the following estimate for these factors ratio $K  \simeq \Lambda_{GP}\Lambda_G$. 
As seen  from Fig.~3, $K$ markedly  exceeds unity, particularly at $\eta \ll 1$.

\begin{figure}[t]
\centering 
\includegraphics[height=5.0cm]{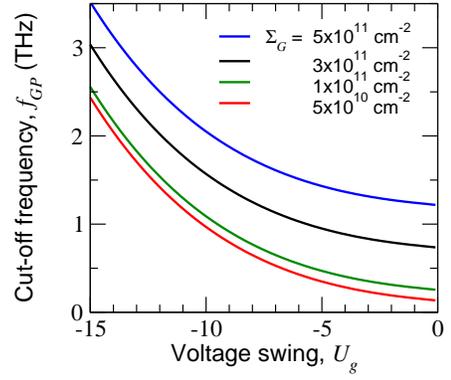}
\caption{The cut-off frequency $f_{GP}$ versus the voltage swing $U_g$ for different scatterer densities $\Sigma_G$.}
\label{F5}
\end{figure}

\section{Detectivity of the GP-bolometers}


The dark-current-limited detectivity of the GP-bolometers $D_{GP}^*$ can be evaluated as (see, for example,~\cite{44}:
\begin{equation}\label{eq29}
D_{GP}^* = \frac{R_{GP}}{\sqrt{4eJ_0H/A}}.
\end{equation}
where $J_0H$ is the net source-drain dc current

Equations~(18) and (24) for relatively low frequencies$\Omega$ yield

\begin{equation}\label{eq30}
D_{GP}^* = D_0^* \frac{|\Lambda_{GP}|}{\sqrt{(1 + P_N)}(1 + 6|U_g|/\pi^2)}
\end{equation}
with

\begin{equation}\label{eq31}
D_0^* = \frac{6\alpha}{\pi^{3/2}\sqrt{\kappa}}\frac{v_W^2\tau_0^{3/2}\tau_0^{\varepsilon}}{T_0}\sqrt{\frac{eE_{SD}}{LT_0}} \propto \tau_0^{3/2}.
\end{equation}
for an arbitrary $E_{SD}$, and

\begin{equation}\label{eq31}
{\rm max}D_0^* = \frac{6\alpha}{\pi\sqrt{\kappa}}\frac{v_W^{3/2}\tau_0^{5/4}(\tau_0^{\varepsilon})^{3/4}}{T_0L^{1/2}}
\end{equation}
for $E_{SD} = \overline{E^*}_{SD}$.
Using  the same parameters as  for the above estimates of  max~$R_{0}^V$ and assuming that $L = 10^{-3}$~cm,
at $E_{SD} = \overline{E^*}_{GP}$, one obtains $D_0^* \simeq (0.122 -2.17)\times 10^{9})$~cm$\sqrt{Hz}$/W.

The GP-bolometer dark-current-limited detectivity $D_{GP}^*$, given by Eq.~(30), exhibits a fairly steep drop with increasing $U_g$ resembling that of the responsivity $R_{GP}^V$ shown in Fig.~3. One should note that
the difference in the  detectivities for a smaller $\Sigma_G$ and those corresponding to a larger $\Sigma_G$
is more pronounced that the pertinent difference in the responsivities.  


\section{Discussion}

\subsection{General comments}

As seen from Eq.~(11), the quantity $\Lambda_{GB} = \displaystyle \frac{1}{\sigma_0} \frac{d \Sigma_{GP}}{d \ln T)}\biggr|_{T = T_0}$
determines the variation of the GP-channel conductivity due to the carrier heating. Its absolute value 
$|\Lambda_{GP}|$ exhibits a maximum at a certain value of $U_g$, which depends on $\Sigma_G$. This is seen from Fig.~3.
Considering only  the variation of the channel conductivity  associated with the carrier transfer to the P-layer, from Eq.~(10)
we find  for $\Lambda_{GP}$ and the  maximum of its modulus $|\Lambda_{GP}|^{max}$

\begin{equation}\label{eq33}
\Lambda_{GP} \simeq -\frac{P_N}{(1 + P_N)^2}\biggl(1 + \frac{\Delta}{T_0}\biggr),
\end{equation}

\begin{equation}\label{eq34}
 |\Lambda_{GP}|^{max} \simeq \frac{1}{4}\biggl(1 + \frac{\Delta}{T_0}\biggr),
\end{equation}
respectively.
This maximum is  reached at $P_N = 1$. i.e., at relatively moderate population of the P-layer ($\Sigma_P = \Sigma_G$) that corresponds to $\mu_0 \simeq  \Delta - T_0\ln(\Sigma_N/\Sigma_G)$
or $U_g = - [\Delta/T_0 -  \ln(\Sigma_N/\Sigma_G)]^2 = U_g^{max}$. For $\Sigma_G = 5\times 10^{10}$, $1\times 10^{11}$, and 
$5\times 10^{11}$~cm$^{-2}$, the pertinent values of $U_g$ are approximately equal to -6, -10, and -23, respectively. The latter
is in line with plots in Fig.~2. It is interesting that max $|\Lambda_{GP}|^{max}$ and $|U_g|$ 
are a linear and a quadratic functions of
$\Delta$, respectively.

The obtained results show that the responsivity and detectivity of the GP-bolometers steeply decrease with increasing
voltage swing $U_g$ (see Fig.~3). 
Thus, it is practical to use the range of relatively small $|U_g|$, although the bandwidth of the GP-bolometers
extends   with increasing $|U_g|$ as seen from Fig.~5. Thus, there is an opportunity of the voltage control 
of the cut-off frequency.  In the GP-LDs, the minimum value of $|U_g|$
 is determined  by the acceptor density  in the GP-channel, so that this density should be minimized to achieve 
 acceptable characteristics. Apart from easier fabrication,  the GP-LDs can exhibit the enhance performance of the whole bolometric photodetector due to
a more effective THz radiation input.


As demonstrated above, the characteristics of   the GL-LD and GL-FET bolometers under consideration can be markedly 
different depending on the Auger parameter $\eta$. This parameter depends on the substrate material, particularly, on its dielectric constant $\kappa$. 
The calculations~\cite{29} predicted the optical phonon recombination time in the G-layers from less than a picosecond to several picoseconds at the carrier densities
under consideration above  and room temperature.
The experimental results of the carrier recombination dynamics in G-layer~\cite{45} were interpreted assuming that
the interband relaxation is associated with the optical phonon processes rather than the carrier-carrier processes,
so that $\tau_{Opt} < \tau_{Auger}$ (and $\eta$ is close to unity). 
The  recent calculations~\cite{29} 
(as well as the previous one's~\cite{46})  
showed  that an increase in $\kappa$
leads to a virtually linear increase in $\tau_{Auger}$ and, hence, in an increase in $\eta$. 
Although, this increase in $\tau_{Auger}$ is not too pronounced - the change in $\kappa$
from 5 to 25 results in a fourfold rise of $\tau_{Auger}$ at room temperature~\cite{29}.  
For example, in the case of GP-LDs with SiO$_2$ and hBN substrates, in which $\kappa \sim 4-5$, $\tau_{Auger} \lesssim 1$~ps, whereas bin the case of the HfO$_2$ substrate, $\tau_{Auger} \gtrsim 2$~ps.
In the GP-FETs, the screening of the carrier interaction by a highly conducting gate can substantially suppress the Auger processes. Indeed, using the data obtained recently~\cite{29}, one can find that $\tau_{Auger}$ being $\tau_{Auger} \simeq 1$~ps at $\kappa = 5$ and the gate layer thickness $W_g = 10-15$~nm,
becomes $\tau_{Auger} \simeq 6$~ps at $W_g = 2$~nm.
One needs to point out that in the case of high-$\kappa$ substrates, additional recombination channel associated with the substrate polar phonons~\cite{47} can promote father increase in $\eta$.
Setting $\tau_{Opt} = (1 - 3)$~ps, we find that the latter values of $\tau_{Auger}$ correspond to $\eta = 0.25 - 0.85$.
Hence, the range of the Auger parameter $\eta$ variations  assumed in the above  calculations appear to be reasonable.


The  values of the GP-detector responsivity demonstrated in Fig.~3  are of the same order of magnitude or can exceed  the room temperature responsivity of the proposed and realized THz photodetectors based on  different 
heterostructures~\cite{21,48,49,50,51,52,53,54,55,56,57,58,59}, including those based on the P-channel~\cite{7,60,61}
(although in G-based devices at very low temperatures much higher responsivities have been achieved~\cite{20}). 

\subsection{Assumptions}

The main assumptions of our device model are fairly natural and practical.
 We disregarded the contribution of the carriers in the P-layer to the net conductivity of the G-P-channel. 
The pertinent condition can be presented as $\sigma_0  \gtrsim eb_P\Sigma_N\exp[(\Delta - \mu_0)/T_0]$ or $\tau_0 \gtrsim \pi\hbar^2\Sigma_N\exp[(\Delta - \mu_0)/T_0]b_P/eT_0$
where $b_P$ is the carrier mobility in the P-layer. Assuming that $b_P = 330 -540$~cm$^2$/s V~\cite{60} (see also~\cite{2})
 $\Sigma_N = 1.2\times 10^{13}$~cm$^{-2}$, and $\mu_0 < 100$~meV (i.e., $|U_g| < 25 $, see Fig.~2),the above inequalities are valid if $\tau_0 \gg 0.005$~ps. 
The values of $\tau_0$ assumed in our calculations well satisfy this requirement.

(i) Above we estimated the scattering time $\tau_p$ and, hence, $\tau_0$ as in~\cite{15}:
$\tau_p^{-1} = v_Sp/\hbar$, so that $\tau_0^{-1} = (v_S/v_W)(T_0/\hbar)$, where
$v_S = (\pi^2U_S^2l_S^2\Sigma_G/4\hbar^2v_W)$,  $U_S = e^2/\kappa\,l_S$ is the characteristic potential of the scatter, and $l_S$ is the screening length.
Setting $l_S = 5$~nm or smaller (see also the estimate for $l_S$ at $\mu_0 = 75$~meV~\cite{13} and below), we find $v_S \simeq 2\times 10^{7}$~cm/s. At $\Sigma_G = 10^{12}$~cm$^{-2}$ and $T_0 = 25$~meV, the latter yields $\tau_0 \simeq 0.12$~ps.
For the scattering on the acoustic phonons due to the deformation potential interaction with the longitudinal vibrations at $T_0 = 25$~meV,
one obtains $v_S(ac) \simeq 8\times 10^5$~cm/s and $\tau_0^{ac} \simeq 3$ps.
The contribution of  the hole-hole scattering in the G-layer to its dc and ac conductivity
are small (despite substantially non-parabolic hole spectrum~\cite{59}). The role of the hole-electron scattering is also small due low electron densities, particularly, at high gate voltages.  

(ii) The interband absorption of the incident radiation with the photons with the  energies $\hbar\Omega \leq 2\pi\hbar\,f_{GP} < 2\mu_0$ in the G-layer disregarded in our model, is practically prohibited due to  the Pauli blocking. 
At small values of $\mu_0$ (i.e., small $|U_g|$), this absorption is weak in comparison with the intraband (Drude) absorption if  

\begin{equation}\label{35}
 \frac {\pi \alpha}{4}  \ll \frac{4\pi\Sigma_{GP}}{c (1 + \Omega^2\tau^2)}. 
\end{equation}
At $\Omega < \tau^{-1} = 2\pi\,f_{GD}$,
inequality (31) implies $\tau_0 > (\pi \hbar/16 T_0) \simeq 0.005$~ps.

(iii)
Considering the features of the DoS (see Fig.~1(e)), the screening length, $l_S$, of the charges in the GP-channel at  low and relatively high  voltage swing  $U_g$ and $T = T_0$
is given by

\begin{equation}\label{36}
l_S \lesssim  \biggl\{\frac{\kappa \hbar^2v_W^2}{8\ln 2 e^2T_0}, \qquad 
l_S \lesssim  \frac{\kappa \hbar^2v_W^2}{4 e^2\mu_0}\biggr\},
\end{equation}
respectively
Assuming $\kappa = 4$ and setting $\mu_0 < T_0 =25$~meV and $\mu_0 = 60$~meV, from   Eq.~(32) we obtain $l_S \simeq 3.4 - 7.5$~nm. The products of the characteristic carrier wavenumbers $k_T = T_0/\hbar\,v_W$ and $k_{\mu_0} = \mu_0/4\hbar\,v_W$ and the pertinent values of $l_S$ are $k_Tl_S \lesssim 0.31$ and  $k_{\mu}l_S \lesssim 0.43$. 
The latter indicates  a rather short range interaction (an effective screening of the charged impurities and the heavy carriers in the P-layer) in the device under consideration at its working conditions.

\section{Conclusions}
We studied the effect of THz photoconductivity 
 of the G- and GP-channels and showed that their conductivity decreases under the THz irradiation (the effect of  negative conductivity). It was revealed that this effect in G-channels  
is determined by the competition of the interband  transitions associated with optical phonons and the Auger generation-recombination processes and vanishes when the latter processes prevail. However, the negative conductivity in the GP-channels is weakly sensitive
to the relative roles of the latter process. 
The negative photoconductivity 
 of the ungated and gated GP-channels (GP-LDs and GP-FETs) under the THz irradiation,
enables using these devices 
 as bolometric  THz photodetectors.
We evaluated the responsivity, bandwidth,  and detectivity characteristics of such THz bolometers and 
 demonstrated that an effective transfer of the carriers from the G-layer into the P-layer, caused by
their heating due to the intraband absorption of the THz radiation, leads to  the substantial decrease in the G-P-channel conductivity. This effect of the negative THz photoconductivity is associated primarily with the intensification of the light carrier scattering in the G-layer on the heavy carriers in the P-layer.
Using the developed device model for the GP-LD and GP-FET bolometers, we demonstrated that these photodetectors can exhibit 
a fairly high responsivity in a wide range of the THz frequencies at the room temperature. The main requirement 
to achieve the elevated photodetector performance is having sufficiently high values of the G-layer mobility.
The main characteristics of the GP-FET bolometers are effectively controlled by the gate voltage.  
The GP-LD and GP-FET THz bolometric photodetectors can substantially surpass the  THz bolometers with the G-channel and  compete and even outperform the existing devices. Further enhancement of the GP-LD and GP-FET THz bolometer can be realized using 
the  GP-GP-...-GP superlattice heterostructures as the channel,  integrating the GP-LDs and GP-FETs with  THz microcavities or waveguides, 
  and  implementing different schemes of the plasmonic enhancement of the THz absorption.

\section*{Appendix A. Short-range versus long-range scattering}
\setcounter{equation}{0}
\renewcommand{\theequation} {A\arabic{equation}}

As seen from Fig.~2,  in the  G-channels the quantity $\Lambda_G < 0$.
This implies that the carrier heating in the G-layers by the absorbed radiation leads to a decrease in the conductivity,
i.e. to the negative photoconductivity. This phenomenon was observed in the experiments (see, for example,~\cite{18,19,37,38}).
 As shown above, the G-layer negative photoconductivity at the room temperatures can appear when the short-range scattering dominates and the Auger generation-recombination processes are weaker than those associated with the optical phonons. In the case of the dominant long-range scattering, the G-layer conductivity rises with increasing carrier effective temperature~\cite{15}. In the model~\cite{18}, the G-layer conductivity was considered assuming that $\tau_p = \tau_0$ is independent of carrier momentum $p$. In such a model, 

\begin{eqnarray}\label{A1}
\sigma_G \propto \int_0^{\infty}d\xi\xi\frac{d(f^e + f^h)}{d\xi}\nonumber\\
 = \ln\biggl[\biggl(1 + e^{\mu^e/T}\biggr)\biggl(1 + e^{\mu^h/T}\biggr)\biggr].
\end{eqnarray}
At small and high ratios $\mu^e/T$ and $\mu^h/T$, Eq.~(A1) yields $\sigma_G \propto [4 + (\mu^e + \mu^h)/2T]$
and  $\sigma_G \propto  (\mu^e + \mu^h)/T]$, respectively.  This results (accounting for Eq.~(4)) in
$\sigma_G \propto [4 + \eta\hbar\omega(1/T - 1/T_0]$ and  $\sigma_G \propto \eta\hbar\omega(1/T - 1/T_0)$.
One can see that at $\tau_p = const$, as in the case  $\tau_p \propto 1/p$ considered  by us, the G-layer conductivity
decreases with increasing carrier temperature. However, this effect vanishes (the conductivity becomes insensitive to the carrier temperature variation and, hence, to the irradiation) when the Auger parameter $\eta$ tends to zero.
Thus, even in this case, the carrier temperature dependence of the GP-channel conductivity $\sigma_{GP}$, related to $\sigma_G $ according to Eq.~(10),
corresponds to the negative photoconductivity with the main contribution of the carrier transfer to the P-layer. 

On the contrary, if the  long-range scattering with $\tau_p \propto p$ would  dominate, $\sigma_G$ could be a rising function of  the carrier temperature leading to the positive G-layer photoconductivity. This can surpass the effect of the G-to-P carrier transitions. 
Both types of the G-layer photoconductivity (negative and positive)  depending on the photon energy and the enviromental gases have been observed, for example, in ~\cite{38}.  
 
\section*{Appendix B. Frequency dependence of the GP- and G-channel conductivity}
\setcounter{equation}{0}
\renewcommand{\theequation} {B\arabic{equation}}

Following the standard procedures (see, for example,~\cite{32}), the ac conductivity can be presented as (compare with
Eq.~(6))

\begin{equation}\label{eqB1}
\sigma_{GP,\Omega} = -\frac{\sigma_0\Sigma_G}{\Sigma_G + \Sigma_P} \int_0^{\infty}\frac{d\xi\,[d(f^e + f^h)/d\xi]}
{1 + \displaystyle\frac{\Omega^2\tau_0^2\Sigma_G^2}{(\Sigma_G + \Sigma_P)^2\xi^2}}.
\end{equation}
At low and high frequencies frequencies, Eq.~(B1) can be rewritten as 

\begin{equation}\label{eqB2}
\sigma_{GP,\Omega} \simeq \sigma_{GP}\simeq\frac{\sigma_0\Sigma_G}{(\Sigma_G + \Sigma_P)},
\end{equation} 

\begin{equation}\label{eqB3}
\sigma_{GP,\Omega} \simeq \frac{\sigma_{GP}}{\Omega^2\tau^2} \simeq\frac{\sigma_0\Sigma_G}{(\Sigma_G + \Sigma_P)}\frac{1}{\Omega^2\tau^2},
\end{equation} 
respectively,
where 

\begin{equation}\label{eqB4}
\frac{1}{\tau} = \frac{2}{\tau_0}\frac{(\Sigma_G + \Sigma_P)}{\Sigma_G}\biggl[{\cal F}_1\biggl(-\frac{\mu_0}{T_0}\biggr) 
+ {\cal F}_1\biggl(\frac{\mu_0}{T_0}\biggr)\biggr]^{1/2}. 
\end{equation}
For $U_g \simeq 0$ and $U_g >\gg  1$, Eq.~(B4) yields

\begin{equation}
\tau \simeq \frac{\sqrt{3}\tau_0}{\pi(1 + P_N)} \qquad \tau \simeq \frac{1}{(1 + P_N)\sqrt{2|U_g|}},
\end{equation}\label{eqB5}
so that $\tau$ as a function of $|U_g|$ can, for example,  be interpolated by Eq.~(16).
In the case of  the dominating long-range scattering, for the cut-off frequency
one obtains $f_{GP} \simeq 1/2\pi\tau_0$.

\section*{Acknowledgments} The authors are grateful to P. P. Maltsev, A. Satou, D. Svintsov, and V. Vyurkov for useful discussions. VR is also thankful to N. Ryabova for assistance.he work  was supported by  Japan Society for Promotion of Science, KAKENHI Grant No. 16H06361, the Russian Science Foundation (Grant No.14-29-00277), Russian Foundation for Basic Research (Grant No. 18-07-01145), RIEC Nation-Wide Collaborative Research Project, and by Office of Naval Research (Project Monitor Dr. Paul Maki).


\begin{thebibliography}{99}

\bibitem{1}
A. H. Castro Neto, F. Guinea, N. M. R. Peres, K. S. Novoselov, and A. K. Geim, 
\lq\lq The electronic properties of graphene,\rq\rq 
Rev. Mod. Phys. {\bf 81}~109--162(2009).


\bibitem{2} 
Xi Ling, H. Wang, S. Huang, F. Xia, and M. S. Dresselhaus,
\lq\lq The renaissance of black phosphorus,\rq\rq
Proc. Nat. Acad. Sci USA, {\bf 112}, 4523--4530 (2015).




\bibitem{3}
F. Bonaccorso, Z. Sun, T. Hasan, and A. Ferrari,
\lq\lq Graphene photonics and optoelectronics,\rq\rq
Nat. Photonics {\bf 4}, 611--622 (2010).

\bibitem{4}
V. Ryzhii, M. Ryzhii, V. Mitin, and T. Otsuji, 
\lq\lq Toward the creation of terahertz graphene injection laser,\rq\rq 
J. Appl. Phys. {\bf 110}, 094503 (2011).


\bibitem{5}
Q. Bao and K. P. Loh, 
\lq\lq Graphene photonics, plasmonics, and broadband optoelectronic devices,\rq\rq
ACS  Nano {\bf 6}, 3677--3677 (2012).

\bibitem{6}
A. Tredicucci and M. Vitiello, 
\lq\lq Device concepts for graphene-based terahertz photonics,\rq\rq 
J. Sel. Top. Quant. {\bf 20}, 130--138 (2014).





\bibitem{7}
M. Buscema, D. J. Groenendijk, S. I. Blanter, G. A. Steele,  H. S. J. van der Zant, and A. Castellanos-Gomez,
\lq\lq Fast and broadband photoresponse of few-layer black phosphorus field-effect transistors,\rq\rq
Nano Lett. {\bf 14}, 3347--3352 (2014).

\bibitem{8}
M. Engel, M. Steiner, and Ph. Avouris,
\lq\lq A black phosphorus photo-detector for multispectral high-resolution imaging,\rq\rq
Nano Lett. {\bf 14}, 6414--6417 (2014).

\bibitem{9} 
E. Leong, R. J. Suess, A. B. Sushkov, H. D. Drew, T. E. Murphy, and M. Mittendorff, \lq\lq Terahertz photoresponse of black phopsporus,\rq\rq
Opt. Express {\bf 25}, No.~11, 12666--12674 (2017).


\bibitem{10}
F. Ahmed, Y. D. Kim, M. S. Choi, X. Liu, D. Qu, Z. Yang, J. Hu, I. P. Herman, J. Hone, W. J. Yoo, 
\lq\lq High electric field carrier transport and power dissipation in multilayer black phosphorus field effect transistor with dielectric engineering,\rq\rq
Adv. Funct. Mater. {\bf 27}, 1604025 (2017).

\bibitem{11}
Y. Deng, Z. Luo, N. J. Conrad, H. Liu, Y. Gong, S. Najmaei, P. M. Ajayan, J. Lou, X. Xu, P. D. Ye, 
\lq\lq Black phosphorus-monolayer MoS$_2$ van der Waals heterojunction p-n diode,\rq\rq
ACS Nano {\bf 8}, 8292--8299 (2014).

\bibitem{12} F. H. L. Koppens, T. Mueller, Ph. Avouris, A, C. Ferrari, M. S. Vitiello, and M. Polini,
\lq\lq Photodetectors based on graphene, other two-dimensional materials and hybrid systems,\rq\rq
Nat. Nanotech. {\bf 9}, 780--793 (2014).

\bibitem{13}
V. Ryzhii, M. Ryzhii, D. Svitsov, V. Leiman, P. P. Maltsev, D. S. Ponomarev, V. Mitin, M. S. Shur, and T. Otsuji,
\lq\lq Real-space-transfer mechanism of negative differential conductivity in gated graphene-phosphorene hybrid structures: Phenomenological heating model,\rq\rq J. Appl. Phys. {\bf 124 } (2018), in press
[arXiv: 1806.06227 (2018)].


\bibitem{14}
Y. Cai, G. Zhang, and Y.-W. Zhang,
\lq\lq Layer-dependent band alignment and work function of few-layer phosphorene,\rq\rq
Sci. Reports {\bf 4}, 6677 (2014).



\bibitem{15}
F. T. Vasko and V. Ryzhii, 
\lq\lq Voltage and temperature dependencies of conductivity in gated graphene,\rq\rq 
Phys. Rev. B {\bf 76}, 233404 (2007).

\bibitem{16}
O. G. Balev, F. T. Vasko, and V. Ryzhii, \lq\lq Carrier heating in intrinsic graphene by s strong dc electric field\rq\rq
 Phys. Rev. B{\bf 79}, 165432 (2009).

\bibitem{17}
V. Ryzhii, T Otsuji, M. Ryzhii, N. Ryabova, S. O. Yurchenko, V. Mitin,
and M. S. Shur, \lq\lq Graphene terahertz uncooled bolometers,\rq\rq
J. Phys. D: Appl. Phys. {\bf 46}, 065102   (2013).


\bibitem{18}
J. N. Heyman, J. D. Stein, Z. S. Kaminski, A. R. Banman, A. M. Massari, and J. T. Robinson,
\lq\lq Carrier heating and negative photoconductivity in graphene,\rq\rq
J. Appl. Phys. {\bf 117}, 015101 (2015).

\bibitem{19}
Xu Du, D. E. Prober, H. Vora, and C. Mckitterick,
\lq\lq Graphene-based bolometers,\rq\rq
Graphene 2D Mater. {\bf 1}, 1--22 (2014).



\bibitem{20} 
Qi Han, T. Gao, R. Zhang, Yi Chen, J. Chen, G. Liu, Y. Zhang, Z. Liu,
X. Wu, and D. Yu,
\lq\lq Highly sensitive hot electron bolometer
based on disordered graphene,\rq\rq
Sci Rep. {\bf  3},   3533 (2013). 


\bibitem{21}
G. Skoblin, J. Sun, and A. Yurgens,
\lq\lq Graphene bolometer with thermoelectric readout and capacitive coupling to an antenna,\rq\rq
Appl. Phys. Lett. {\bf 112}, 063501 (2018)

\bibitem{22}
G. Zhang, A. Chaves, S. Huang F. Wang1, Q. Xing, T. Low, and H. Yan1,
\lq\lq Determination of layer-dependent exciton binding energies in few-layer black phosphorus,\rq\rq
Sci. Advances  {\bf 16} Mar 2018:
Vol. 4, no. 3, eaap9977.

     
\bibitem{23} 
T. Low, R. Roldán, H. Wang, F. Xia, P. Avouris, L. M. Moreno, F. Guinea, 
\lq\lq Plasmons and screening in monolayer and multilayer black phosphorus,\rq\rq 
Phys. Rev. Lett. {\bf 113}, 106802 (2014).




\bibitem{24}
S. Yuan,
A. N. Rudenko, and M. I. Katsnelson, 
\lq\lq Transport and optical properties of single- and bilayer black phosphorus with defects,\rq\rq
Phys. Rev. B {\bf 91}, 115436 (2015).

\bibitem{25}
 J. Xi, M. Long, D. Wang, and Z. Shuai, 
\lq\lq First principles prediction of charge mobility in carbon and organic nanomaterials,\rq\rq 
 Nanoscale {\bf 4}, 4348--4369 (2012).



\bibitem{26}
F. Rana, P. A. George , J. H. Strait, S. Sharavaraman, M. Charasheyhar, and M. G. Spencer, 
\lq\lq Carrier recombination and generation rates for intravalley and intervalley phonon scattering in graphene,\rq\rq
Phys. Rev. B {\bf 79}, 115447 (2009).


\bibitem{27}
V. Ryzhii, M. Ryzhii, V. Mitin, A. Satou, and T. Otsuji, 
\lq\lq Effect of heating and cooling of photogenerated electron-hole plasma in optically pumped graphene on population inversion,\rq\rq
Jpn. J. Appl. Phys. {\bf 50}, 094001 (2011).








\bibitem{28}
M. S. Foster and I. L. Aleiner, 
\lq\lq Slow imbalance relaxation and thermoelectric transport in graphene,\rq\rq
Phys. Rev. B {\bf 79}, 085415 (2009).

\bibitem{29} 
G. Alymov, V. Vyurkov, V. Ryzhii, A. Satou, and D. Svintsov,
\lq\lq Auger recombination in Dirac materials: A tangle of many-body effects,\rq\rq
Phys. Rev. B  {\bf 97}, 205411 (2018).





\bibitem{30}
J. S. Blakemore, {\it Semiconductor Statistics}, Dover, 1987.



\bibitem{31}
 T. Ando, 
\lq\lq Screening Eefect and impurity scattering in monolayer graphene,\rq\rq 
 J.  Phys, Soc. Japan {\bf 75}, 074716 (2006)

\bibitem{32}
L. A. Falkovsky and A. A. Varlamov, 
\lq\lq Space-time dispersion of graphene conductivity,\rq\rq
European Phys. J. B {\bf 56}, 281--284 (2007).

\bibitem{33}
E. H. Hwang, S. Adam, and S. D. Sarma, 
\lq\lq Carrier transport in two-dimensional graphene layers,\rq\rq
Phys. Rev. Lett. {\bf 98}, 186806 (2007).


\bibitem{34}
 V. Vyurkov and V. Ryzhii, 
 \lq\lq Effect of Coulomb scattering on graphene conductivity,\rq\rq 
JETP Lett. {\bf 88}, 370--372 (2008).



\bibitem{35}
E. H. Hwang and  S. Das Sarma, 
\lq\lq Acoustic phonon scattering limited carrier mobility in two-dimensional extrinsic graphene,\rq\rq
Phys. Rev. B
{\bf 77},  115449 (2008).



\bibitem{36}
E. H. Hwang and  S. Das Sarma, 
\lq\lq Screening induced temperature dependent transport in 2D graphene,\rq\rq
Phys. Rev. B {\bf 79}, 165404 (2009).


 
\bibitem{37}
G. Jnawali, Y. Rao, H. G. Yan, and T. F. Heinz,\lq\lq Observation of a transient decrease in terahertz conductivity of single-layer graphene induced by ultrafast optical excitation,\rq\rq 
Nano Lett. {\bf 13}, 524--530 (2013).



\bibitem{38}
C. J. Docherty, C. T. Lin, H. J. Joyce, R. J. Nicholas, L. M. Hertz, L. J. Li, and M. B. Johnston,
\lq\lq Extreme sensitivity of graphene photoconductivity to environmental gases,\rq\rq
Nat. Comm. {\bf 3},1228 (2012)

 
 



\bibitem{39}
S. V. Morozov, K. S. Novoselov, M. I. Katsnelson, F. Schedin, D. C. Elias, J. A. Jaszczak, and A. K. Geim, 
\lq\lq Giant intrinsic carrier mobilities in graphene and its bilayer,\rq\rq
Phys. Rev. Lett.{\bf 100}, 016602 (2008).



\bibitem{40}
H.  Hirai, H. Tsuchiya, Y. Kamakura, N. Mori, and M. Ogawa,
\lq\lq Electron mobility calculation for graphene on substrates,\rq\rq
J. Appl. Phys. {\bf 116}, 083703 (2014).

 
\bibitem{41}
L. Banszerus, M. Schmitz, S. Engels, J. Dauber, M. Oellers, F. Haupt, K. Watanabe, T. Taniguchi, B. Beschoten, and C. Stampfer, \lq\lq Ultrahigh-mobility graphene devices from chemical vapor 
deposition on reusable copper,\rq\rq Science Advances {\bf 1}, No. 6. , e1500222 (2015). 


\bibitem{42}
   L. Wang, I. Meric, P. Y. Huang, Q. Gao, Y. Gao, H. Tran, T. Taniguchi, K. Watanabe, L. M. Campos, 
   D. A. Muller, J. Guo, P. Kim, J. Hone, K. L. Shepard, and C. R. Dean,\lq\lq One-dimensional electrical contact to a two-dimensional material,\rq\rq
Science {\bf 342},  614-617 (2013).


 
 \bibitem{43}
K. J. Tielrooij, J. C.W. Song, S. A. Jensen, A. Centeno, A. Pesquaera, A. Z. Elorza, M. Bonn, L. S. Levitov, and F. H.L. Koppens, \lq\lq Photoexcitation cascade and multiple hot-carrier generation in graphene,\rq\rq
Nat. Phys. {\bf 9}, 248--252~(2013).



 \bibitem{44}
 H. Schneider and H,C, Liu, {it Quantum  Well Infrared Photodetectors: Physics and Applications}, Springer, NY, 2007. 


\bibitem{45}
J. M. Dawlaty, S. Shivaraman, M. Chandrashekhar, F. Rana, and M. G. Spencer 
\lq\lq Measurement of ultrafast carrier dynamics in epitaxial graphene,\rq\rq
Appl. Phys. Lett. {\bf 92}, 042116 (2008).

\bibitem{46}
F. Rana, \lq\lq Electron-hole generation and recombination rates for coulomb scattering
in graphene, \rq\rq 
Phys. Rev. B {\bf 76}, 155431 (2007).




\bibitem{47}
F. Rana, J. H. Strait, H, Wang, and  C, Manolatou,
\lq\lq Ultrafast carrier recombination and generation rates for plasmon emission and absorption in graphene,\rq\rq
 Phys. Rev. B {\bf 84}, 045437 (2011). 





\bibitem{48}
S. D. Gunapala, S. V. Bandara, J. K. Liu, J. M. Mumolo, S. B. Rafol, D. Z. Ting, A. Soibel, and C. Hill, \lq\lq Quantum Well Infrared Photodetector Technology and Applications,\rq\rq	
IEEE J. Sel. Topics Quant. Electron. {\bf 20}, No. 6 (2014). 

\bibitem{49}
 V. Ryzhii, T. Otsuji, V. E. Karasik, M.Ryzhii, V. Leiman, V. Mitin, and M. S. Shur, \lq\lq Comparison of intersubband quantum-well and interband graphene layer infrared photodetectors,\rq\rq IEEE J. Quant. Electron. {\bf 54}, No. 2 (2018).
 
\bibitem{50} 
 V. Ryzhii, M. Ryzhii, M. S. Shur, V. Mitin, A Satou, and T Otsuji,
\lq\lq Resonant plasmonic terahertz detection  in graphene split-gate field-effect transistors with lateral p–n junctions,\rq\rq J. Phys. D: Appl. Phys. {\bf 49},  315103 (2016).


\bibitem{51}
M. Mittendorff, S. Winnerl, J. Kamann, J. Eroms, D. Weiss, H, Schneider1, and M. Helm,
\lq\lq Ultrafast graphene-based broadband THz detector,\rq\rq Appl. Phys. Lett. {\bf 103}, 021113 (2013).

\bibitem{52}
 V.Ryzhii, M.Ryzhii, D.Svintsov, V.Leiman, V.Mitin, M.S.Shur,and T .Otsuji,\lq\lq Nonlinear response of infrared photodetectors based on van der Waals heterostructures with graphene layers,\rq\rq Optics Express {\bf 25},5536-–5549
(2017). 

\bibitem{53}. 
V. Ryzhii, M. Ryzhii, V. Leiman, V. Mitin, M. S. Shur, and T. Otsuji, \lq\lq Effect of doping on the characteristics of infrared photodetectors based on van der Waals heterostructures with multiple graphene layers,\rq\rq
J. Appl.Phys. {\bf 122}, 054505 (2017). 

\bibitem{54}
V. Ya. Aleshkin, A. A. Dubinov, S. V. Morozov, M. Ryzhii, T. Otsuji, V. Mitin, M. S. Shur, and V. Ryzhii,
\lq\lq Interband infrared photodetectors based on HgTe-CdHgTe quantum-well heterostructures,\rq\rq
Opt. Mat. Exp. {\bf 8}, 1349 (2018). 

\bibitem{55}
V. Ryzhii, M. Ryzhii, V. Mitin, and T. Otsuji, \lq\lq Terahertz and infrared photodetection using p-i-n multiple-graphene-layer structures,\rq\rq
J. Appl. Phys. {\bf 107}, 054512 (2010).

\bibitem{56}
A. V. Muraviev, S. L. Rumyantsev, G. Liu, A. A. Balandin, W. Knap, and M. S. Shur,
\lq\lq Plasmonic and bolometric terahertz detection by graphene field-effect transistor,\rq\rq 
Appl. Phys. Lett. {\bf 103}, 181114 (2013)

\bibitem{57}
Y. Wang, W. Yin, Q. Han, X. Yang, H. Ye, Q. Lv, D. Yin, \lq\lq Bolometric effect in a waveguide-integrated graphene photodetector,\rq\rq Chin Phys. B {\bf 25}, 118103 (2016). 

\bibitem{58} D. A. Bandurin, D. Svintsov, I. Gayduchenko, S. G. Xu, A. Principi, M. Moskotin, I. Tretyakov, D. Yagodkin, S. Zhukov, T. Taniguchi, K. Watanabe, I. V. Grigorieva, M. Polini, G. Goltsman, A. K. Geim, and G. Fedorov,
\lq\lq Resonant terahertz detection using graphene plasmons,\rq\rq 
arXiv: 1807.04703 (2018).

\bibitem{59}
D. S. Ponomarev, D. V. Lavrukhin, A. E. Yachmenev, R. A. Khabibullin, I. E. Semenikhin, V. V. Vyurkov, M. Ryzhii, T. Otsuji, and V. Ryzhii1, \lq\lq Lateral terahertz hot-electron bolometer based on an array of Sn nanothreads in GaAs,\rq\rq
J. Phys. D: Appl. Phys. {\bf 51}, 135101  (2018).
 



\bibitem{60}
L. Viti, J. Hu, D. Coquillat, A. Politano, W. Knap, and   M. S. Vitiello,
\lq\lq Efficient Terahertz detection in black-phosphorus nano-transistors with selective and controllable plasma-wave, bolometric and thermoelectric response,\rq\rq Scientific Reports {\bf 6},  20474 (2016).   
%

\bibitem{61}
E. Leong, R. J. Suess, A. B. Sushkov, H. D. Drew, T. E. Murphy, and M. Mittendorff,
\lq\lq Terahertz photoresponse of black phosphorus,\rq\rq
Optics Express {\bf 25}, 12666--12674 (2017).

\bibitem{62}
D. Svintsov, V. Ryzhii, A. Satou, T. Otsuji, and V. Vyurkov, \lq\lq Carrier-carrier scattering and negative dynamic conductivity in pumped graphene,\rq\rq
Optics Express {\bf 22 }, 19873--19886 (2014).







 




\end{thebibliography}
\end{document}